# NUCLEAR ASTROPHYSICS BEFORE 1957

Edwin E. Salpeter, Space Sciences
Cornell University, Ithaca, NY 14853

Abstract: I discuss especially my summer with Willy Fowler at Kellogg Radiation in 1951, where I did my "triple-alpha" work. I also go back even earlier to Arthur Eddington and Hans Bethe. I also mention the 1953 summer school in Ann Arbor

This Caltech. meeting celebrates the 50 year anniversary of $B^2FH$, the famous paper by Burbidge, Burbidge, Fowler and Hoyle (1), plus a similar but shorter paper by Al Cameron (2). Of course, most of the meeting is concerned with developments in nuclear astrophysics since 1957, but my talk is purely about events before 1957 and, especially, about my stay at Kellogg Radiation Lab in July and August 1951. Thus, of the five people being celebrated I concentrate on Willy Fowler, my temporary boss during that summer. However, I will also mention Hans Bethe, the discoverer of the C, N, O - cycle twelve years earlier, and even the influence (both good and bad) of Arthur Eddington.

It is interesting to look for the effect which a scientist's Ph.D. advisor and other role models had on their subsequent work. For Willy Fowler (1911-1995) this is pretty simple: He got his Ph.D. with Charlie Lauritsen in 1936, they both continued at Kellogg Lab for a lifetime and Willy considered

Charlie as his main mentor in all aspects of experimental nuclear physics. Nuclear Theory he had to absorb from others, but it was part of Willy's genius that he could run the main center for nuclear astrophysics with equal emphasis on theory and experiment. Hans Bethe (1906-2005), on the other hand, had two contrasting role models: (I) Arnold Sommerfeld, the advisor for his Ph.D. thesis in 1928, who emphasized rigorous mathematics and carrying theoretical calculations to completion; (ii) Enrico Fermi, his boss in 1931-32, who emphasized using intuition, taking shortcuts and making physics fun. To his "youngsters" at Cornell (including myself) Bethe advised "learn rigorous mathematics in case you need it, but in any particular problem use only the minimum necessary". In my own case I had three role models, my thesis advisor Rudi Peierls, then Hans Bethe and then Willy Fowler for nuclear problems.

I started working with Hans Bethe at Cornell's Lab of Nuclear Studies in 1949, mainly in Quantum Electrodynamics which was very much in vogue then – partly because of a rivalry between Feynman at Cornell and Schwinger at Harvard. However, both Peierls and Bethe worked in astoundingly many scientific fields and some of that rubbed off onto me. Following my academic grandfather Fermi more than Sommerfeld, I have also worked in many fields. Without Fermi's genius, intuition does not always work but I have enjoyed the variety in spite of some of my blunders. At any rate, I was also working a little in nuclear theory during 1949 and 50. At that time Bethe was not working directly in Astrophysics, but was giving Willy Fowler extensive advice on the relevant nuclear theory. Bethe's help extended as far as "lending Willy one of his young men", so I was sent to Kellogg Rad. Lab. for the summer of 1951. I will return to this summer later on, but first some earlier history, starting with Arthur Eddington's book (3), written in 1924/25.

The stellar models in Eddington's book had the merit that they were simple enough to give explicit results for the sun and other stars, but some of the assumptions were too simple and too single-minded. In particular, he stipulated that there is no convection anywhere in a stellar interior and no chemical inhomogenetics, not even for Red Giant stars. He was honest enough to admit in the book's last paragraph that "Somewhere in the present tangle of evolution and sources of energy I have been misled; and my guidance of the reader must terminate with the admission that I have lost my way". In spite of this frankness Eddington never changed his mind in the ten or more years after 1925 (he died in 1944), even though others set the record straight.

Eddington's combination of brilliant insight with stubborn intransigence is illustrated by his views on hydrogen: He knew $E=mc^2$ very well and, as soon as accurate masses became known for hydrogen and helium in 1920, he pointed out forcefully that converting four hydrogen atoms into one helium atom would release a lot of energy in a star's interior. However, he did not pursue this idea for stellar models because he assumed that hydrogen was an almost negligible constituent of stars (less than 10% by mass). He persisted in this extreme assumption long after Cecilia Payne-Gaposchkin (1900-79) had shown, in her brilliant Harvard Ph.D. thesis in 1925, that hydrogen is by far the most abundant element in stars.

Payne-Gaposchkin is interesting in many ways (apart from having been Jesse Greenstein's thesis advisor). One of these is detailed by Olivia Walling (in History of Science at U. C. Santa Barbara) in a 2005 dissertation, mainly about Willy Fowler (4). She points out the extreme importance of a director or other leader encouraging a brilliant female scientist in an institution that is otherwise extremely male chauvinist. This happened to Payne-Gaposchkin at Harvard, but Caltech. was similar if not more so in the 1940's and 50's. How Fowler and Greenstein encouraged and helped Margaret Burbidge, and in turn benefitted from her productivity, may well crop up later at this meeting. Walling also mentions the similar case of Fay Ajzenberg-Selove who first came to Kellogg in 1952 and had a fruitful collaboration with Tommy Lauritsen (Charlie's son) on compiling energy levels of light nuclei.

During the 1930's the new quantum mechanics was being applied to nuclear theory of relevance to astrophysics, starting with some joint Bethe-Peierls papers on the deuteron. Of great importance was the "Bethe Bible", a massive review article on everything that was known about nuclear dynamics (5). Also important was George Gamow's presentation of the formula for Coulomb barrier penetration which led to general formulae for thermonuclear reaction rates. The possibility of a proton-proton collision leading to the formation of a deuteron had already been suggested by others, but at the very beginning of 1938 Gamow suggested to his graduate student Critchfield to collaborate with Bethe on an explicit calculation.

Bethe and Critchfield had a fairly accurate expression for the deuterium production rate as a function of density and temperature by February 1938. They were a bit vague on the completion of the proton-proton chain from deuterium to helium, but this completion is so fast it did not matter for the energy production rate. However, the density and temperature and hydrogen abundance at the center of the sun mattered a lot and they still used the values from Eddington's 1925 models. With the incorrect hydrogen abundance used by Eddington, the density and temperature were also quite inaccurate and the Bethe-Critchfield formulae gave a badly wrong value for the solar luminosity. Bethe considered their work "just an exercise" and was ready to give up astrophysics for good.

Fortunately Gamow organized a conference in Washington on "Energy Production in Stars" for March 1938. Although reluctant to go, Bethe finally attended and heard from Bengt Stroemgren that Eddington's values were badly wrong. Stroemgren presented reliable temperature/density values for main sequence stars of different masses,using Payne-Gaposchkin's fully accepted hydrogen abundance. With these reliable stellar models the proton-proton chain would now give good values for the luminosity of the sun and of less massive main sequence stars (6).

Bethe soon noticed that the proton-proton chain did NOT work for main sequence stars appreciably more massive than the sun, where the observed luminosity increases much more rapidly with increasing temperature than the calculations gave. He therefore knew that there must be an additional mechanism for converting hydrogen into helium, involving higher nuclear Coulomb charges where the thermonuclear reaction rates would be more temperature sensitive. No, he did not solve the problem on his train-ride back to Ithaca (that actually did happen for his "low-brow" calculation of the Lamb shift, many years later). Back at Cornell he looked at all the possibilities and it took him two whole weeks to discover the carbon-nitrogen cycle (now called the CNO-cycle) in April 1938. He was also pleased by the fact that, starting with carbon you mostly did not end up with oxygen but ended up with carbon plus helium, so that carbon and nitrogen act mainly as catalysts. He was even more pleased by the fact that the sum of the proton-proton chain and CNO-cycle worked fine for main sequence stars of all masses (7). In this paper he also noted that this did not work at all for Red Giant stars. That is where my trip to Pasadena twelve years later comes in.

Soon after the hydrogen-burning papers, Bethe joined the Manhattan Project full time during World War II and had to leave in abeyance two questions: (I) The completion of the proton-proton chain and (ii) what provides energy production in red giant stars. When he sent me to Kellogg Lab and Willy Fowler in 1951 these two questions were of obvious interest, but I expected to look at other things also. Apart from energy production in stars, Herman Bondi, Tommy Gold and Fred Hoyle had raised another controversy with their Steady State Cosmology: In this cosmology the universe was always the same as now, there was no Big Bang with its high temperature and all the medium and heavy elements had to be made in stars. George Gamow was a particularly forceful proponent of the opposite point of view that all element production, starting from hydrogen, took place during the big Bang. Many of us, including myself, were rooting for the Steady State (maybe an early form of opposition to creationism?).

Pasadena and Caltech. at the beginning of July 1951 were a great revelation for me, both scientifically (I had never learnt any astronomy before that) and socially. Socially, Cornell was a friendly place but somewhat subdued, so I was not prepared for Caltech's exuberance. In our first week Willy and Ardie Fowler invited us to a 5:30 p.m. Cocktail Party. We knew that such parties do not last longer than two hours, so we put a roast in the oven before we left. Needless to say, the party went on all night and we all spent the early morning skinny-dipping in someone elses pool (I can't remember whose). We were almost arrested but Willy and Ardie (she was quiet, but effectively forceful) placated the police and all ended happily except that our roast was burned.

My education in nuclear experimentation and in astronomy also began with a bang. Besides the two stalwarts of Willy and Charlie Lauritsen I spent time with Ward Whaling, Bob Christy and Jesse Greenstein at Kellogg and met the whirlwind astronomer Walter Baade. Especially important for me was a visit of Martin Schwarzschild from Princeton University, who had started to make stellar models numerically with early versions of powerful computers. In particular he was studying the evolution of red giant stars where the build-up of helium in the core drives a chemical inhomogeneity, with the outer layers expanding and the core contracting and heating up. His work was not yet published but he was already able to tell me that central temperatures somewhere between 1 and 2 times $10^8$K are reached at the tip of the red giant branch.

I am proud of the fact that I was able to carry out two different lengthy calculations, both in July of that year (8, 9). In both cases the most time-consuming effort was trying out various alternatives before finding the correct one. The major achievement there was not really mine but the groundwork Bethe had provided in his detailed publications (plus Fermi's admonition to run through preliminaries with a light touch). One of my two papers was on the completion of the proton-proton chain and more accurate values for its energy production rate. I did this work mainly for the sake of completeness and it did not excite much interest at the time, although it was of some relevance much later on for the solar neutrino problem (8). The other calculation, on nuclear reactions beyond helium and energy production at the tip of the red giant branch, was of immediate interest and got me tenure at Cornell just a few years later (9).

Nuclear reactions starting with $He^4$ presented a problem, both for stars and for the Big Bang, because of the "gaps" at atomic mass 5 and at 8. Since neither $He^4 + H^1$ nor $He^4 + He^4$ leads to a stable nucleus, some kind of three-body reaction was needed but that left many possibilities open. For instance, I investigated $2He^4 + H^1$ with little luck first and finally returned to the "triple-alpha reaction" which had been suggested before. In fact Öpick had written a paper on this before me (10), but it was unknown to me and others till much later. He also was not aware of the resonance in $Be^8$ and got much too small a reaction rate. I was lucky being in Pasadena at the time, because Willy Fowler was able to tell me that $Be^8$, although not stable, was "almost stable" with a resonance energy of only about 95 keV. With this resonance at an advantageous energy, not only was the energy production rate much larger than without it, but the calculation was actually simpler. My original handwriting in Fig. 1 shows that only one page was needed to derive the formula.

However, this was not the end of the triple-alpha story: Although there were some earlier suggestions of a resonance level in $C^{12}$ at an appropriate energy, Willy Fowler in 1951 considered these suggestions as wrong. As a consequence my $C^{12}$ – formation paper (9) was done without one, but I mentioned that a resonance might raise the rate by a factor of about 1000. I needed almost $2 \times 10^8$ K, slightly larger than the Schwarzschild estimate in 1952 for the tip of the red giant branch, but within the range of the uncertainty. However, the further absorption of alpha-particles to form $O^{16}$ (and then $Ne^{20}$) would give quite a large O/C abundance ratio. This ratio was known to be close to unity instead of large and I should have "smelled a rat", but did not. Fred Hoyle, on the other hand, did so just a few years later and showed that a $C^{12}$ resonance was needed and predicted at what energy it should appear (11). Fowler, et al at Kellogg Lab were skeptical of this theoretical prediction but nevertheless carried out new experiments. They found a $C^{12}$ resonance level just about where Hoyle had predicted (12) which started a long collaboration between Fowler and Hoyle. At some stage of the Big Bang expansion the same temperature is reached as in a red giant star, but at a very much lower density. Three-body reaction rates are very sensitive to density and a negligible amount of carbon would be produced.

After my two successful calculations in July, I spent most of August 1951 calculating what reactions would be important at temperatures between (1 and 2) x $10^9$ K. There was already a paper by Hoyle (13), on energy loss reactions (leading to pre-supernova collapse) at about (4 or 5) x $10^9$ K and I wanted to avoid those complications. I filled many more pages in August than I had in July and Fig. 2 shows a page of just one of the sections. I put some of all this in my detailed lecture notes for the Ann Arbor summer school in May 1953, but I never published this work. In the light of the data available then, the calculations were not even wrong. I just could not put a coherent story together, whereas $B^2FH$ did only a few years later.

I have spent so much time on unpublished work to emphasize that we should discuss at meetings not only successes but also failures and that failures are often due to a lack of "follow-through". Hans Bethe has given examples of follow-through: After military work leading to nuclear weapons and Nagasaki during World War II and soon after, he went on to work for test-ban treaties and disarmament instead and more recently went on to make a speech (1995) that scientists should now "cease and desist from further weapons work".  His increasing pessimism stemmed partly from him having lived through the Weimar Republic, i.e. Germany just BEFORE Hitler and before World War II.  He noted that intellectuals then were not necessarily against democracy nor for war, but just did not take the time to get involved in politics.  What worried him particularly in his last few years in Ithaca was the increasing apathy of U. S. intellectuals in the face of increasing threats to civil rights – reminiscent of the Weimar Republic.

I did not witness the Weimar Republic myself, but I also feel strongly that U. S. scientists should take more time out from their own work to speak out on public issues.  This is especially true for topics where we have some technical background (even if not direct expertise) such as environmental remedies for global warming or for depleted uranium.  A few of us might even get involved in part-time research toward improving alternative energy sources, for instance, but we all have the right background to note the technical pitfalls in some government proposals.  Two glaring examples are Ballistic Missile Defense and deep penetrating Bunker Busters.  The Union of Concerned Scientists (www.ucsusa.org), an organization funded by private donations, puts out reports on these and similar topics from time to time.  The Center for Constitutional Rights (www.ccr-ny.org) is similarly active on legal matters.  I feel scientists should even speak out occasionally on political matters where they do not have expertise, only objectivity.  The disasters which would follow any kind of U. S. attack on Iran, for instance, are so enormous that we should all protest.

Even apart from military work, we should consider the potential harm which scientists might cause with their own work, even if it sounds beneficial. Martin Rees has pointed out potential dangers especially from nano-technology, computer-advances and genetic engineering, but there must be others as well. Maybe we should start considering some kind of "Hippocratic Oath" to be undertaken by practicing scientists. Such considerations might also point out previously unknown potential dangers.

Almost as important for my astrophysics education as the Caltech. summer of 1951 was the Ann Arbor Astrophysics Summer School in 1953 at U. Michigan. Most important for me was Walter Baade explaining the two stellar populations in detail. Having to prepare my own Ann Arbor lectures was also useful, but the second most important influence was George Gamow, who talked on everything and was most impressive on the Early Universe. I did not follow his example of drinking Vodka at lunchtime from a water pitcher and I also disagreed with him on one other, more scientific, point: It was known already that massive main sequence stars are needed to spread heavy elements through interstellar space during their death. Gamow pointed out, correctly, that few massive stars are visible today and went on to say, incorrectly, that massive stars are not important for heavy elements in interstellar space. Only a small fraction of the massive stars born in the lifetime of our Galaxy are alive today, but I argued that the much more numerous dead ones were the important ones anyway. Straight after the Ann Arbor summer school I spent a year at the newly opened Australian National

University and calculated stellar birth rates from present-day luminosity functions. This led to my paper on the initial mass function (14).

I won't give any history of my IMF work in 1954, since I have a long essay on this (15), but I end with the following footnote: When Fred Hoyle and I shared the Crafoord Prize in 1997 we had to give speeches in front of the King of Sweden. According to the citation, Fred was expected to talk about $B^2FH$ and I was expected to talk about my 1951 triple-alpha calculation. Instead, I talked about my later IMF paper (14) and Fred talked about his earlier paper (13).

(Two figures below, with figure number small at the bottom):

FIGURE CAPTIONS

Fig. 1: One sheet of my original handwritten notes in July 1951 (in blue ink, yet, not black), deriving my triple-alpha rate formula.

Fig. 2: A page of my notes in August 1951 on likely nuclear reactions around 1 or 2 times $10^9$K.

July 1951

# $3\,He^4 \rightarrow C^{12}$

Simple reaction:

$$p\,/\text{gm sec.} = \frac{4}{3^{5/2}}\,\rho\,\frac{x_1 x_2}{m_1 m_2}\,\frac{\Gamma}{\hbar}\,e^{(16R/a)^{1/2}}\,\tau^2 e^{-\tau}$$

$$P = \frac{m_2}{\rho\, x_1 x_2}\,p = ~~\text{prob/sec. for } \ldots~~ . \quad P\rho x_1 = \frac{p\, m_2}{x_2} = \text{prob/sec for one nuc. 2.}$$

$x$ is %age by weight, $m$ = reduced mass

$$\tau = 3\left(\frac{\pi^2 m e^4 Z_1^2 Z_2^2}{2\hbar^2 kT}\right)^{1/3}; \quad a = \hbar^2/me^2 Z_1 Z_2; \quad R = 1.6\times10^{-13}(A_1+A_2)^{1/3}$$

$$P = m_2\; 5.3\times10^{25}\;\Gamma\;\varphi\;\tau^2 e^{-\tau}\;/\text{gm sec.}$$ with $\Gamma$ in eV & $T$ in $10^6\,{}^\circ K$.

$\varphi(\alpha+\alpha) = 1.29$, $\varphi(Be^8+\alpha) = 16.2$

$$\varphi = \frac{1}{A_1 A_2 (Z_1 Z_2 A)^3}\left(\frac{8R}{a}\right)^2 e^{(\;)^{1/2}}$$

<u>A</u> <u>Resonance</u>: $E_r$ = res. energy $= 0.95 \pm 0.10$ MeV (for small density)

$$\sigma = \pi \lambda\!\!\!^{-2}\,\frac{\Gamma_{\alpha\alpha}\Gamma_{Be\alpha}}{(E-E_r)^2 + \frac{1}{4}(\Gamma_{\alpha\alpha}+\Gamma_{Be\alpha})^2}$$

$$\Gamma_{Be\alpha} \ll \Gamma_{\alpha\alpha} \ll E_r$$

$$\bar{P}(Be\alpha) = \frac{\hbar\, m_{Be}}{x_{Be}}\,p(Be\alpha) = \frac{4}{3^{5/2}}\,\rho\,\frac{x_\alpha}{m_\alpha}\,\Gamma_{rad}\,e^{(\;)^{1/2}}\,\tau^2 e^{-\tau}\,aR^2$$

~~$\varphi\, p(3\alpha) = \frac{1}{2}\left(\frac{\rho x_\alpha}{m_\alpha}\right)^2 \frac{2}{\pi^{1/2}}\frac{E_r^{1/2}}{(kT)^{3/2}}e^{-E_r/kT}(2E_r/m_\alpha)^2$~~

$$p(3\alpha) = 10^{-4}(\rho x_\alpha)^2 \left(\frac{x_\alpha}{m_\alpha}\right)\Gamma_{rad}\,\varphi(Be\alpha)\,T^{-3/2}\,e^{-11.7E_r/T}\,\tau^2 e^{-\tau}$$

# $(1-2)\times 10^{9}$°K reactions

Dissociation reactions still unimportant

$2C^{12} \to Na^{23} + p$, $\to Mg^{24}$ } becomes important just below $10^{9}$°K. (At $10^{9}$°K half-life ~ 30 yrs, $\varepsilon \sim 10^{9}$ erg/gm sec).

Mainly $Na^{23}$ produces, liberated protons go through carbon cycle with remaining carbons.

Net effect : $2C^{12} \to Na^{23} + \frac{1}{4}(He^{4} - 2e^{-})$

energy / $2C^{12}$ = $(13.93 - 11.70) + \frac{1}{4}28 + \frac{1}{4}7 \sim 11$ MeV ($\to Mg^{24}$ gives 13.9 MeV)

As $C^{12}$ gets depleted, temp. rises and $[C^{12} + O^{16} \to Si^{28}, Al^{27} + p, Si^{27} + n]$ sets in (mainly $Al^{27}$). p still used up through $C^{12}$ carbon cycle.

When $C^{12}$ completely depleted $[2O^{16} \to S^{32}, P^{31} + p, S^{31} + n, \to P^{31} - e + n]$ sets in. At $2\times 10^{9}$°K halflife ~ 1000 yrs, $\varepsilon \sim 3\times 10^{7}$ erg/gm sec.

p used up through $p + O^{16} \to F^{17} \to O^{17} - e$, $p + O^{17} \to N^{14} + \alpha$

$N^{14}$ goes thro' carbon cycle & gives some $C^{12}$ which combines w. $O^{16}$.

Also $N^{14} + O^{16} \to P^{30*}$ ~~[illegible]~~ $\to P^{30} + 18.3$ MeV $\to Si^{30} + 3.0$ MeV $+ e^{+}$

$\to Si^{29} + p + 12.8$ MeV

$\to P^{29} + n$

Most pble. net reaction : $8O^{16} \to 4P^{31} + 4p$ ; $2p + O^{16} \to N^{14} + \alpha - e$, $N^{14} + 2p \to C^{12} + \alpha - 2e$

$\to O^{16} - 2e$

$4p + O^{16} \to He^{4} + O^{16}$ [illegible]

$\therefore \; 9O^{16} \to 4P^{31} + Ne^{20}$

is page 3 of "Life-times at $>10^{9}$°K", dated Aug. 1951.